\documentclass[aps,prl,twocolumn,superscriptaddress,showpacs]{revtex4-1}
\usepackage{graphics}
\usepackage{subfigure}
\usepackage{longtable}
\usepackage{graphicx}
\usepackage{pstricks}
\usepackage{amsmath}
\usepackage{dcolumn}
\usepackage{nicefrac}
\usepackage{bm}
\newcommand{\barl}{Cu$_4$(OH)$_6$FBr}
\newcommand{\herb}{ZnCu$_3$(OH)$_6$Cl$_2$}

\begin{document}

\title{Barlowite as a canted antiferromagnet: theory and experiment}

\author{Harald O. Jeschke}
\affiliation{Institut f\"ur Theoretische Physik, Goethe-Universit\"at
  Frankfurt, Max-von-Laue-Stra{\ss}e 1, 60438 Frankfurt am Main,
  Germany}

\author{Francesc Salvat-Pujol}
\affiliation{Institut f\"ur Theoretische Physik, Goethe-Universit\"at
  Frankfurt, Max-von-Laue-Stra{\ss}e 1, 60438 Frankfurt am Main,
  Germany}

\author{Elena Gati}
\affiliation{Physikalisches Institut, Goethe-Universit{\"a}t Frankfurt,
  Max-von-Laue-Stra{\ss}e 1, 60438 Frankfurt am Main, Germany}

\author{Nguyen Hieu Hoang}
\affiliation{Physikalisches Institut, Goethe-Universit{\"a}t Frankfurt,
  Max-von-Laue-Stra{\ss}e 1, 60438 Frankfurt am Main, Germany}

\author{Bernd Wolf}
\affiliation{Physikalisches Institut, Goethe-Universit{\"a}t Frankfurt,
  Max-von-Laue-Stra{\ss}e 1, 60438 Frankfurt am Main, Germany}

\author{Michael Lang}
\affiliation{Physikalisches Institut, Goethe-Universit{\"a}t Frankfurt,
  Max-von-Laue-Stra{\ss}e 1, 60438 Frankfurt am Main, Germany}

\author{John A. Schlueter}
\affiliation{Division of Materials Research, National Science
  Foundation, Arlington, Virginia 22230, USA}

\author{Roser Valent{\'\i}}
\affiliation{Institut f\"ur Theoretische Physik, Goethe-Universit\"at
  Frankfurt, Max-von-Laue-Stra{\ss}e 1, 60438 Frankfurt am Main,
  Germany}

\begin{abstract}
  We investigate the structural, electronic and magnetic properties of
  the newly synthesized mineral barlowite {\barl} which contains
  Cu$^{2+}$ ions in a perfect kagome arrangement. In contrast to the
  spin-liquid candidate herbertsmithite {\herb}, kagome layers in
  barlowite are perfectly aligned due to the different bonding
  environments adopted by F$^-$ and Br$^-$ compared to Cl$^-$. We
  perform density functional theory calculations to obtain the
  Heisenberg Hamiltonian parameters of {\barl} which has a Cu$^{2+}$
  site coupling the kagome layers. The 3D network of exchange
  couplings together with a substantial Dzyaloshinskii-Moriya coupling
  lead to canted antiferromagnetic ordering of this compound at
  $T_N=15$~K as observed by magnetic susceptibility measurements on
  single crystals.
\end{abstract}


\maketitle

\begin{figure}
\centering
\includegraphics[width=\columnwidth]{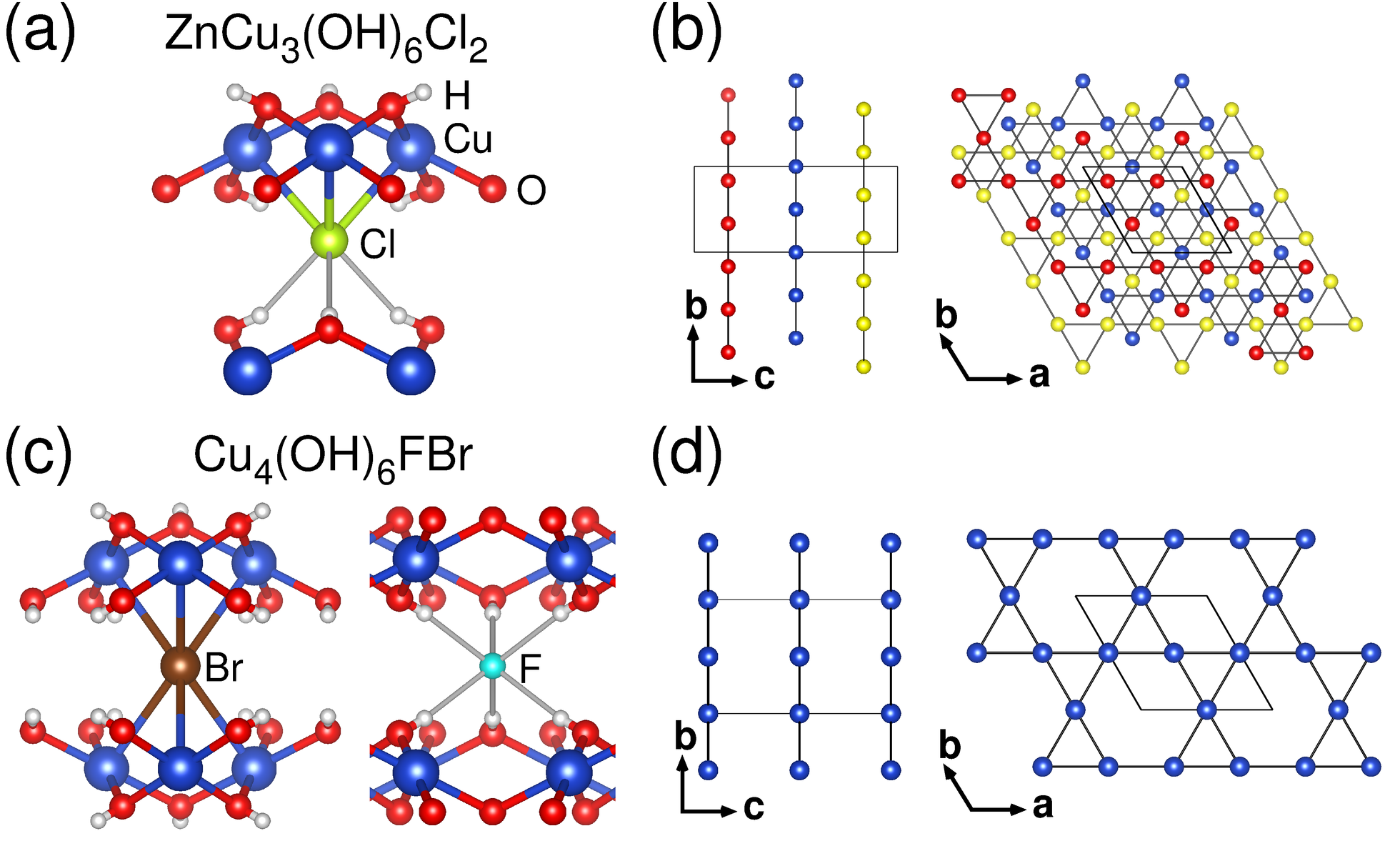}
\caption{Chloride environment of herbertsmithite (a) compared to
  bromide and fluoride environments of barlowite (c). (b) Mix of
  covalent and hydrogen bonding in {\herb} leads to three kagome
  layers shifted with respect to each other, as shown by differently
  colored Cu$^{2+}$ sites. (d) Preference of Br$^-$ for covalent
  bonding and of F$^-$ for hydrogen bonding leads to perfect alignment
  of kagome layers in {\barl} (only Cu(1) sites are shown here). }
\label{fig:environment}
\end{figure}

\begin{figure}
\centering
\includegraphics[width=0.8\columnwidth]{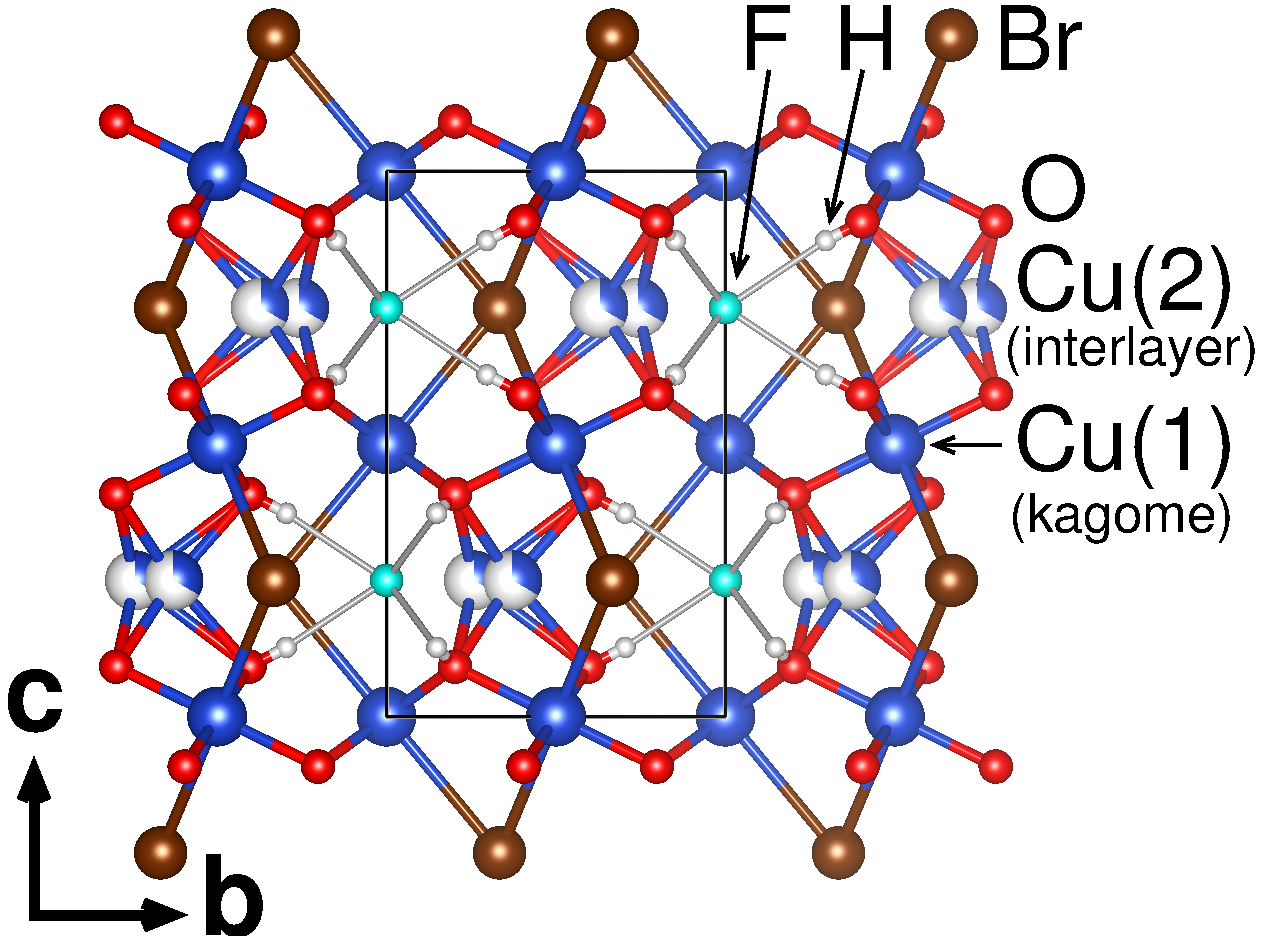}
\caption{Crystal structure of {\barl} ($P\,6_3/mmc$ space group,
  No. 194). Note that the Cu(2) site with Wyckoff position $12j$ is
  $\nicefrac{1}{3}$ filled. }
\label{fig:structure}
\end{figure}

Since the successful synthesis of herbertsmithite
(\herb)~\cite{Shores2005}, spin-liquid candidates based on spin-1/2
kagome lattices have been intensively investigated in recent
years~\cite{Mendels2010,Mendels2011}. The paratacamite family of
compounds has proven to be a fertile ground for kagome materials with
different properties. The Zn$_x$Cu$_{4-x}$(OD)$_6$Cl$_2$ series of
materials has been found to form valence-bond solids due to
distortions of the kagome layer in monoclinic space
groups~\cite{Lee2007}. The replacement of Zn$^{2+}$ in {\herb} by a
nonmagnetic ion like Mg$^{2+}$ possibly leads to another spin-liquid
candidate~\cite{Colman2011}, as does replacement of 2Cl$^-$ by
SO$_4^{2-}$~\cite{Li2014}. On the other hand, magnetic ions between
the kagome layers lead to compounds that order
magnetically~\cite{Li2013}.  While realizations of quantum spin
liquids have been desperately searched for~\cite{Lee2008}, so far,
together with the triangular-lattice molecule-based materials
$\kappa$-(BEDT-TTF)$_2$Cu$_2$(CN)$_3$ (BEDT-TTF =
bis(ethylenedithio)-tetrathiafulvalene)~\cite{Shimizu2003}, and
EtMe$_3$Sb[Pd(dmit)$_2$]$_2$ (dmit =
1,3-dithiole-2-thione-4,5-dithiolate, Me = CH$_3$, Et =
C$_2$H$_5$)~\cite{Itou2008}, herbertsmithite has been considered one
of the best candidates~\cite{Han2012}. However, {\herb} has not been
free from controversy as Zn$^{2+}$ and Cu$^{2+}$ are similar in size,
and both kagome layers diluted with nonmagnetic
Zn$^{2+}$~\cite{Mendels2011} as well as magnetic Cu$^{2+}$ ions
replacing Zn$^{2+}$ between the kagome layers~\cite{Freedman2010} are
possible. Therefore, it would be desirable to devise a crystal
modification of herbertsmithite that would make the Cu$^{2+}$ and
Zn$^{2+}$ sites less similar in order to increase the tendency of
Cu$^{2+}$ ions to form the kagome layer as well as the tendency of the
nonmagnetic transition metal ion to stay away from the kagome
plane. The purpose of this work is (1) to suggest such a strategy,
exemplarily realized in the mineral barlowite and (2) to unveil the
microscopic origin of the electronic and magnetic behavior of this
system via a combination of density functional theory calculations and
magnetic susceptibility measurements.  Besides the search for ideal
quantum spin liquids, the electronic structure of kagome materials
offers a large variety of interesting properties, like for instance
the presence of flat bands at certain fillings that induces Nagaoka
ferromagnetism~\cite{Hanisch1997} or the presence of Dirac points at
4/3 filling that could lead to unusual symmetry protected metals or
superconductors~\cite{Mazin2014}.

In herbertsmithite, the Cl$^-$ binding environment is partially
covalent, partially hydrogen bonded, as shown in
Fig.~\ref{fig:environment}~(a). This leads to a horizontal staggering
of kagome layers (Fig.~\ref{fig:environment}~(b)) as Cu$^{2+}$ triangles
can be placed either above or below a Cl$^-$ ion, but not both above
and below. We suggest using a mixed halide system where the strong
hydrogen bond acceptance of the F$^-$ ion is used to create a hydrogen
rich pocket with six hydroxyl ions; on the other hand, Br$^-$ can form
six covalent bonds to three Cu$^{2+}$ ions above and three below
(Fig.~\ref{fig:environment}~(c)). Following this recipe via the
chemical synthesis of {\barl}, we arrive at perfectly aligned kagome
planes as shown in Fig.~\ref{fig:environment}~(d). This compound is
known as the mineral barlowite~\cite{Elliott2010}. So far, only some
of its properties have been reported~\cite{Han2014}.

Single crystals of barlowite, {\barl}, were grown synthetically
through the hydrothermal reaction of copper carbonate basic
(malachite), with perbromic acid in the presence of ammonium
fluoride. The crystal structure of {\barl} was determined by single
crystal X-ray diffraction measurements~\cite{Xray} at ambient
temperature and is shown in Fig.~\ref{fig:structure}. Barlowite
crystallizes in $P\,6_3/mmc$ symmetry with each intralayer Cu$^{2+}$
(Cu(1)) ion lying on a site of $2/m$ symmetry (see
Table~\ref{tab:struct}). This intralayer Cu$^{2+}$ exhibits a strongly
tetragonally distorted octahedral coordination with four equatorial
Cu-O bonds of 1.954(1)~{\AA} and two axial Cu-Br bonds of
3.022~{\AA}. Interlayer Cu$^{2+}$ (Cu(2)) sites lie on a general
position and are thus 1/3 occupied and disordered over three
equivalent positions.

\begin{table}[hbt]
  \caption{Structural parameters for {\barl} at $T=298(2)$~K  ($P\,6_3/mmc$ space group,
    $a=6.799(4)$~{\AA}, $c=9.3063(13)$~{\AA}, $Z=2$). In this table, $U$ is the anisotropic 
    displacement parameter (isotropic for H).}
\label{tab:struct} 
\begin{tabular}[c]{lllllll}
\hline\hline 
Atom & Site & $x$ & $y$ & $z$ & $U$~({\AA}$^{2}$) & Occ.\\ \hline
Cu(1)&$6g$&  0.5000     & 0& 0& 0.01540(17) & 1\\
Cu(2)&$12j$& 0.62884(12)& 0.2577(2) &$\nicefrac{1}{4}$& 0.0133(4) & $\nicefrac{1}{3}$ \\
O    &$12k$& 0.79768(18)& 0.20232(18) &0.0916(2)& 0.0137(4) & 1\\
H    &$12k$& 0.852(3)   & 0.148(3) &0.127(4)& 0.021 & 1 \\ 
Br   &$2c$&  $\nicefrac{1}{3}$&$\nicefrac{2}{3}$& $\nicefrac{1}{4}$& 0.0184(2) & 1 \\ 
F    &$2b$&  0     &0& $\nicefrac{1}{4}$& 0.0238(10) & 1  \\
\hline\hline
\end{tabular}
\end{table}

\begin{figure}
\centering
\includegraphics[width=\columnwidth]{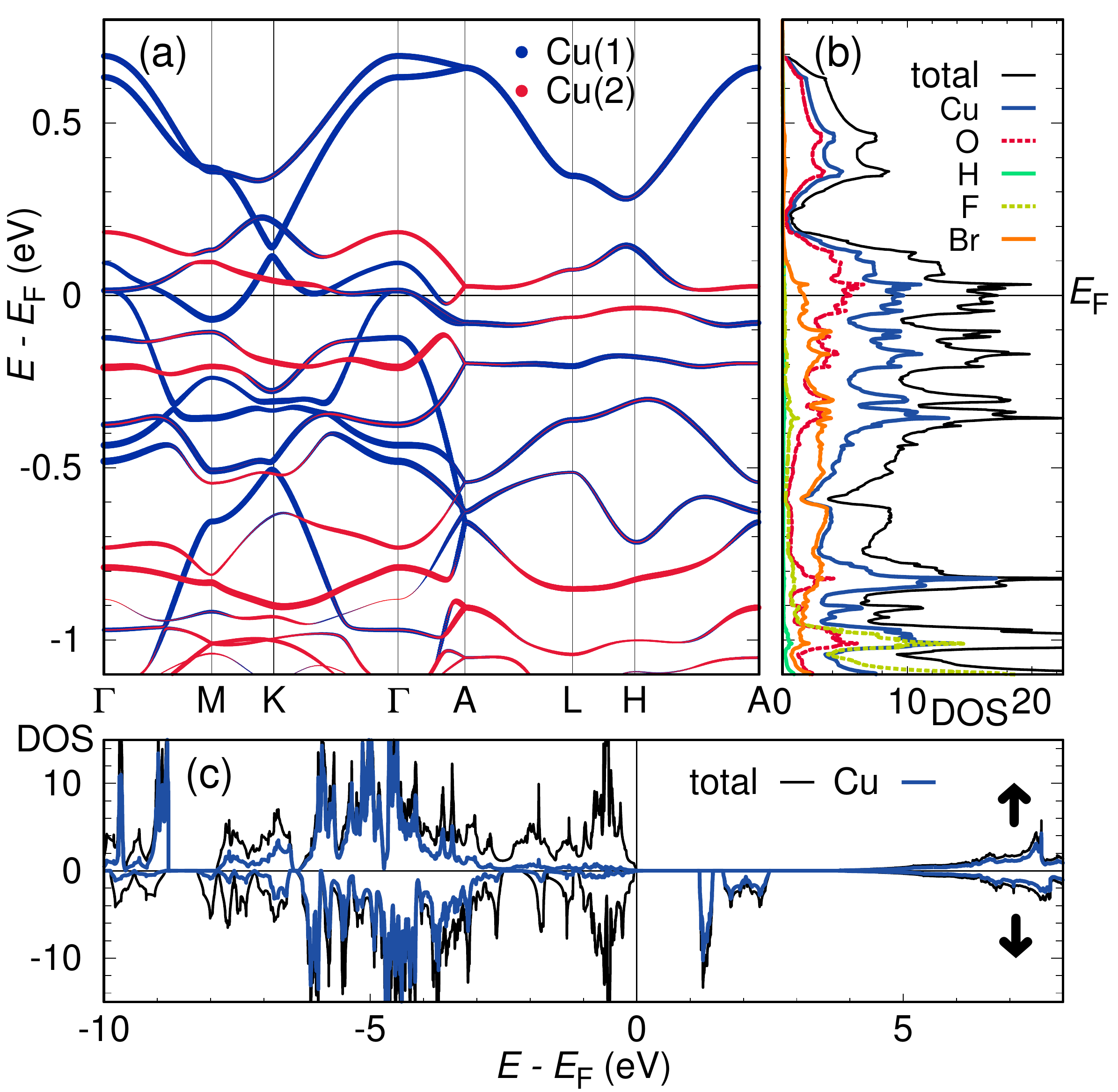}
\caption{(a) GGA band structure and (b) GGA density of states (DOS) of
  {\barl}.  DOS is given in states per eV and formula unit.  High
  symmetry points $M=(\nicefrac{1}{2},0,0)$,
  $K=(\nicefrac{1}{3},\nicefrac{1}{3},0)$, $A=(0,0,\nicefrac{1}{2})$,
  $L=(\nicefrac{1}{2},0,\nicefrac{1}{2})$,
  $H=(\nicefrac{1}{3},\nicefrac{1}{3},\nicefrac{1}{2})$ were chosen to
  reflect the $P\,6_3/mmc$ symmetry of the real material rather than
  the $C\,mcm$ symmetry used for computational purposes. (c) GGA+U
  density of states for $U= 6$~eV, $J_H=1$~eV and ferromagnetic
  order. Note that the size of the gap is related to the value of $U$
  considered.}
\label{fig:bsdos}
\end{figure}

In order to characterize barlowite electronically and magnetically, we
combined first principles density functional theory (DFT) calculations
with susceptibility measurements.  Our DFT calculations were performed
on the basis of the full-potential non-orthogonal local-orbital basis
(FPLO)~\cite{Koepernik1999}, employing the generalized gradient
approximation (GGA)~\cite{Perdew1996} as well as
GGA+U~\cite{Liechtenstein1995} functionals. The Hubbard parameter $U$
was set to 6~eV, the Hunds rule coupling $J_H$ to 1~eV. The
exchange-coupling constants between spin-1/2 Cu$^{2+}$ ions were
obtained by mapping GGA+U total energy differences of several
Cu$^{2+}$ spin configurations onto a spin-1/2 Heisenberg
model~\cite{Foyevtsova2011,Jeschke2013}. In order to make a sufficient
number of Cu$^{2+}$ sites inequivalent to allow for various spin
configurations, the symmetry was lowered from the $C\,mcm$ to the
$C\,m$ space group (No. 8) containing 6 inequivalent Cu$^{2+}$
positions.

In Fig.~\ref{fig:bsdos} (a)-(b) we show first the GGA band structure
and density of states of barlowite.  The main contribution near the
Fermi level is of Cu $3d$ orbitals hybridized with O $2p$
orbitals. The band structure along the high-symmetry path
$\Gamma$-$M$-$K$-$\Gamma$ reflects the dispersion of the kagome layers
(dominated by Cu(1) $d$ states) while the band structure along the
high-symmetry path $A$-$L$-$H$-$A$ at $k_z = 0.5$ arises from the 2D
lattice formed by the inter-kagome Cu(2) atoms.  We observe that the
electronic structure of the kagome layer resembles the electronic
structure of the spin-liquid compound herbertsmithite very
well~\cite{Jeschke2013}. However, both herbertsmithite and barlowite
are Mott insulators.  In order to reflect this behavior in the band
structure calculations we show in Fig.~\ref{fig:bsdos}~(c) the density
of states calculated with the GGA+U functional with $U=6$~eV.

\begin{table}
  \caption{
    Exchange coupling constants for {\barl}, calculated with GGA+U at $U=6$~eV,
    $J_H=1$~eV and with atomic-limit double-counting
    correction.
  }
  \label{tab:J}
\begin{ruledtabular}
  \begin{tabular}{cccr}
    Name & $d_\text{Cu-Cu}$ (\AA) & Type & ~~$J_i(K)$ \\ 
    \hline
    \multicolumn{4}{c}{Kagome layer coupling} \\
    $J_3$ & 3.3399  &  Cu(1)-Cu(1)  & 177\\[0.2cm]
    \multicolumn{4}{c}{Interlayer couplings} \\
    $J_1$ & 2.7632 &  Cu(1)-Cu(2) &-205 \\
    $J_2$ & 3.1885  &  Cu(1)-Cu(2) &-32 \\
    $J_4$ & 4.6532 &  Cu(1)-Cu(1) &5 \\
    $J_6$ & 5.5264 &  Cu(2)-Cu(2) &16
    % $J_8$ & 5.7277 &          &-1.0 
  \end{tabular}
 \end{ruledtabular}
\end{table}

\begin{figure}
\centering
\includegraphics[width=0.8\columnwidth]{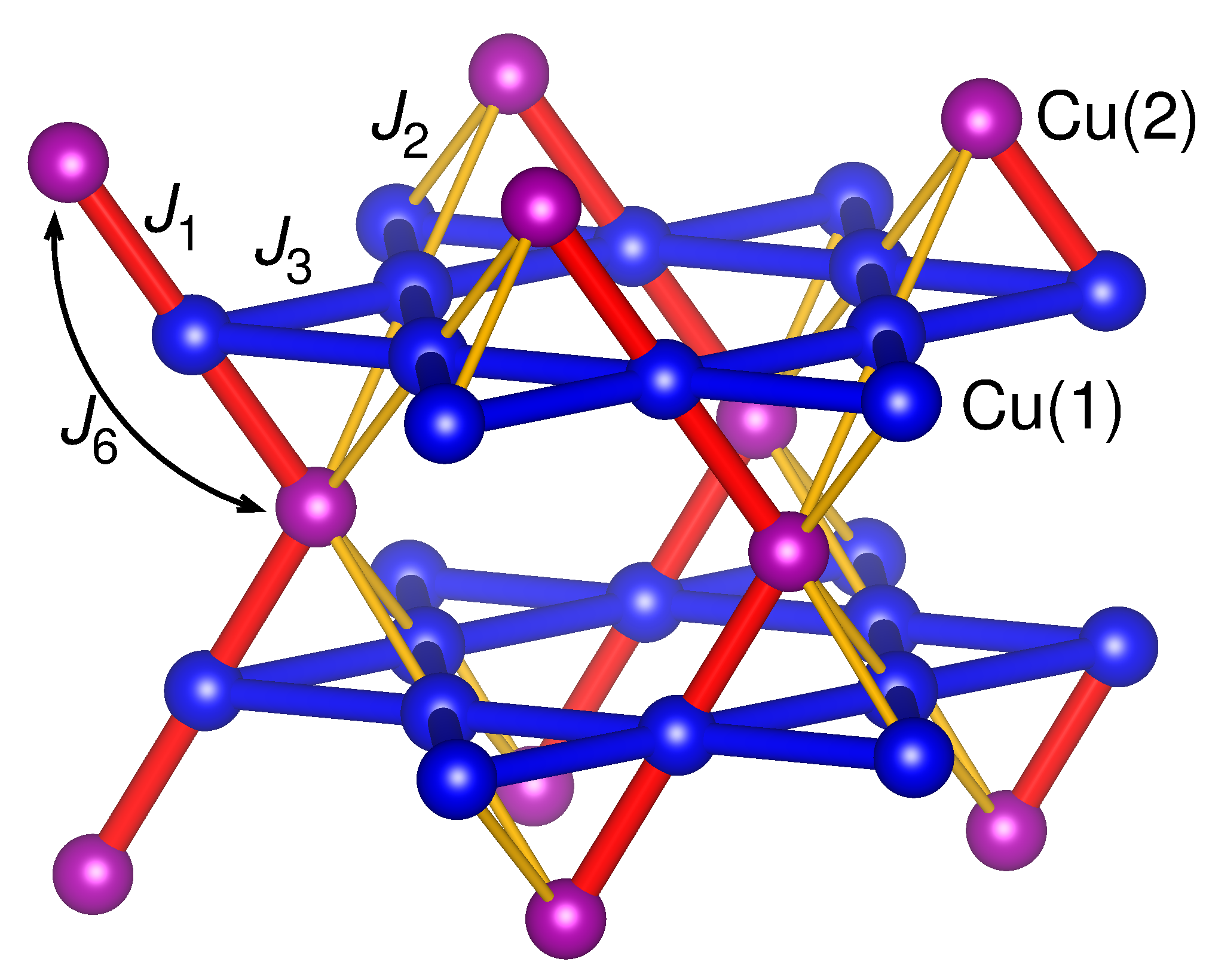}
\caption{Important exchange paths of {\barl}. The notation $J_i$,
  $i=1,2,3,\dots$ denotes 1$^{st}$, 2$^{nd}$, 3$^{rd}$,\dots Cu-Cu
  neighbors.  Positive (negative) $J$ values denote antiferromagnetic
  (ferromagnetic) couplings. Exchange constants were determined to be
  $J_3=177$~K (kagome layer), $J_1=-205$~K and $J_2=-32$~K. A coupling
  $J_6=16$~K connecting two Cu(2) sites along the red bond is
  indicated by a curved arrow. Note that disordered Cu(2) sites in the
  $P\,6_3/mmc$ space group lead to random permutations of the two weak
  and one strong ferromagnetic couplings between every Cu(1) triangle
  above and below a Cu(2) site. Larger couplings are shown with
  cylinders of larger diameter.}
\label{fig:couplings}
\end{figure}

The exchange couplings we obtain from total energy calculations are
listed in Table~\ref{tab:J}.  The nearest neighbor coupling in the
kagome plane is $J_3=177$~K for barlowite. This is very similar to the
value $J=182$~K obtained for herbertsmithite, reflecting the fact that
Cu-O-Cu angles are very similar in both compounds (117$^\circ$ in
barlowite, 119$^\circ$ in herbertsmithite).  The fact that barlowite
has Cu$^{2+}$ ions at interlayer positions (Cu(2)) determines the
magnetic behavior of this system at low temperatures.  Specifically,
we find that ferromagnetic interlayer couplings ($J_1=-205$~K and
$J_2=-32$~K) exist between Cu(1) (in the kagome layer) and Cu(2)
(interlayer). Further exchange paths become increasingly and
significantly weaker -- comparable or smaller than $0.1 J$. Within the
$C\,m$ unit cell chosen here, it was not possible to separate the
coupling $J_5$ corresponding to a Cu(1)-Cu(2) distance of 5.2359~{\AA}
from the small ferromagnetic $J_2$. The vertical coupling $J_4$
between the kagome planes is negligibly small. The resulting
Heisenberg Hamiltonian parameters are illustrated in
Fig.~\ref{fig:couplings}. Note that due to the lowering of the
symmetry from $P\,6_3/mmc$ to $C\,mcm$ for the calculations, the path
of the ferromagnetic one-dimensional Cu(2) chains has become uniquely
defined. In reality, these chains wiggle through the crystal according
to the actual positions of Cu(2) which is randomly chosen from the
three possible sites. We estimate an error bar on the Heisenberg
Hamiltonian parameters of the order of {20\%} and possibly larger for
the smaller couplings because the calculated values depend strongly on
the essentially unknown size of the Hubbard parameter $U$ and because
of the tendency of DFT to overestimate the stability of the
ferromagnetic state.

\begin{figure}
\centering
\includegraphics[width=\columnwidth]{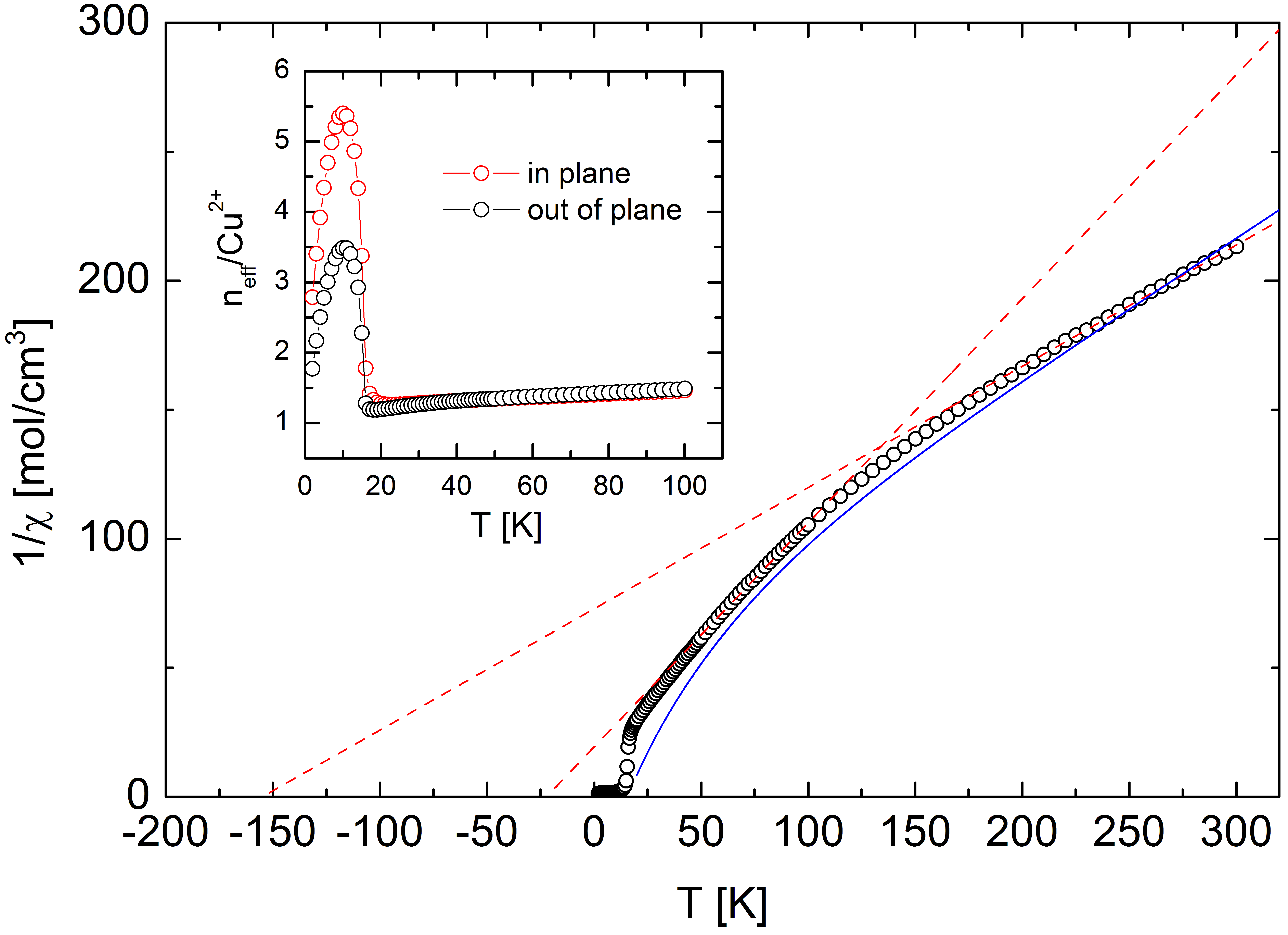}
\caption{Inverse molar susceptibility as a function of temperature
  (open circles) taken at $B = 0.1$~T for $B \parallel c$ together
  with Curie-Weiss fits (red broken lines) for $200~{\rm K} \le T \le
  300~{\rm K}$ and $50~{\rm K} \le T \le 100~{\rm K}$. Also shown is a
  theoretical approximation to $1/\chi$ calculated by 10th-order
  high-temperature series expansion~\cite{Lohmann2014}, using the
  calculated value $J_3=177$~K and $J_1=-0.94J_3$, $J_2=-0.16J_3$ and
  $J_6=0.15J_3$ as well as a $g$ factor of 2.20 (blue solid
  line). Note that $J_1$, $J_2$ and $J_6$ agree with the calculated
  values within their error bars. Inset: Effective magnetic moment per
  Cu$^{2+}$ ion as a function of temperature in fields $B = 0.1$~T for
  $B \parallel c$ and $B \perp c$ both after cooling in zero field
  (ZFC).}
\label{fig:suscept}
\end{figure}

\begin{figure}
\centering
\includegraphics[width=\columnwidth]{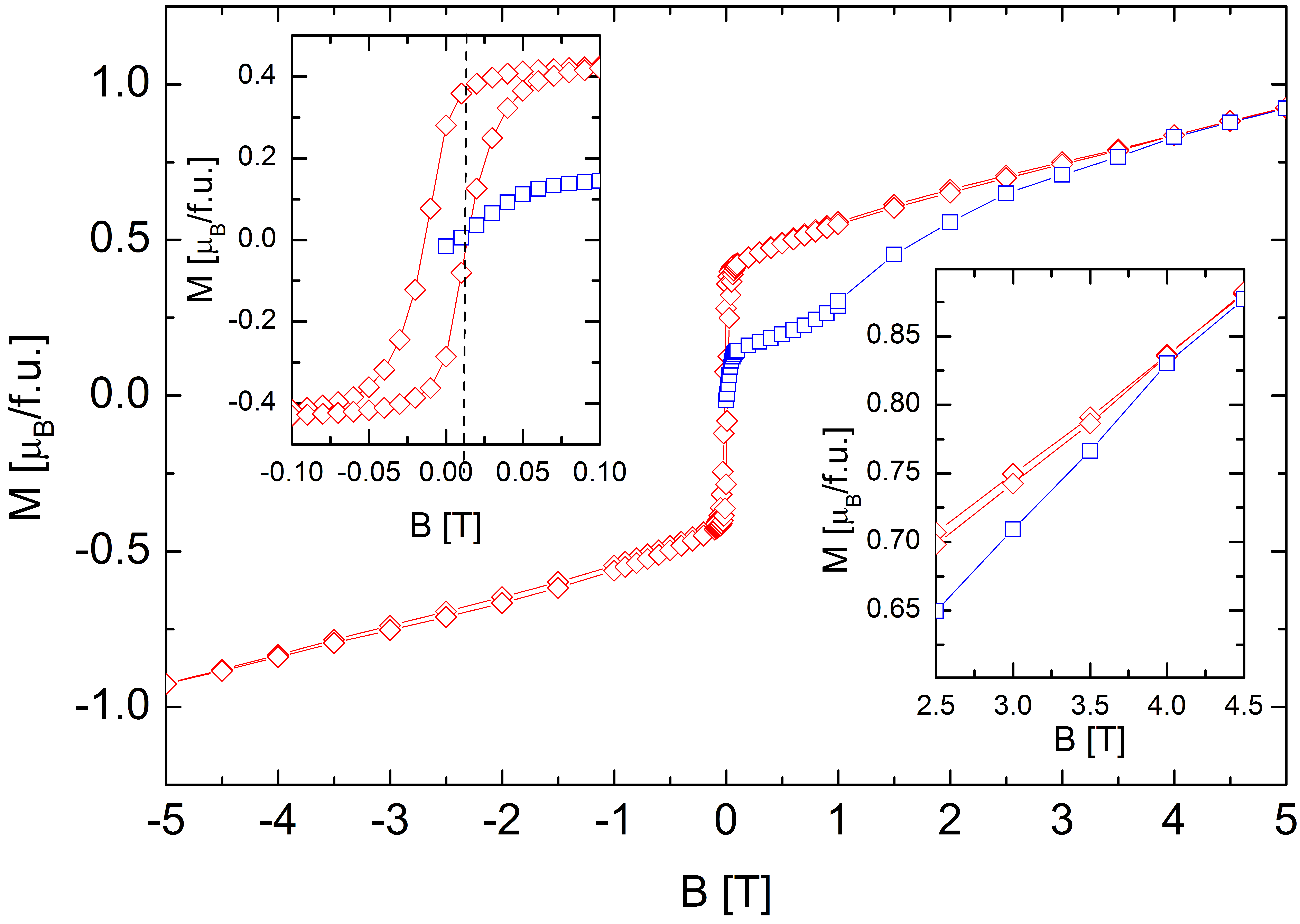}
\caption{Magnetization as a function of magnetic field $B$ measured
  for $B \parallel c$ at $T = 2$~K. The blue open squares correspond
  to the virgin curve taken after cooling in zero field, the red
  diamonds correspond to data taken upon subsequent field
  cycling. Lower right inset: blow-up of high-field section. Upper
  left inset: blow-up of low-field section.  }
\label{fig:magnetization}
\end{figure}

The magnetic properties of barlowite were measured using a commercial
Quantum Design SQUID magnetometer in the temperature range $2~{\rm K}
\le T \le 300~{\rm K}$ on a single crystal of mass $m = 3.4$~mg. The
susceptibility measurements were performed in various fields up to 1~T
with $B$ oriented in plane and along the hexagonal $c$-axis. For the
latter orientation the magnetization was measured in fields up to
$\pm$5~T. The experimental data have been corrected for the
temperature-independent diamagnetic core contribution of the
constituents~\cite{Kahn1993} and the magnetic contribution of the
sample holder.

Figure~\ref{fig:suscept} shows the inverse molar magnetic
susceptibility (open circles) as a function of temperature for
$B \parallel c$. The $1/\chi_{\rm mol}$ data can be approximated by a
Curie-Weiss-like behavior (red broken lines in Fig.~\ref{fig:suscept})
at the upper and lower ends with an antiferromagnetic
Weiss-temperature $\theta = (155 \pm 2)$~K for $200~{\rm K} \le T \le
300~{\rm K}$ and $(22 \pm 2)$~K for $50~{\rm K} \le T \le 100~{\rm
  K}$. This behavior is consistent with the presence of two dominant
magnetic coupling constants of different sign, suggested by our DFT
calculation.

By considering the exchange parameters obtained from DFT, we
calculated $1/\chi_{\rm mol}$ using 10th-order high-temperature series
expansion~\cite{Lohmann2014}.  We find that a very good fit of the
experimental observations, shown by the solid blue line in
Fig.~\ref{fig:suscept}, is obtained with $J_3=177$~K and $J_1 =
-0.94J_3$, $J_2=-0.16J_3$ and $J_6=0.15J_3$ in combination with a $g$
value of 2.20. Given the strongly distorted octahedral Cu$^{2+}$
environment, implying anisotropic $g$ values ranging from $0.1 \le
\Delta g/g \le 0.15$~\cite{Silver1973,Petrashen1980}, a $g$ value of
2.2 for $B \parallel c$ is reasonable.  The values of the exchange
couplings are within the error bars of the DFT calculation confirming
that the DFT analysis of barlowite is reliable.

The inset of Fig.~\ref{fig:suscept} exhibits the effective magnetic
moment $n_{\rm eff}$ per Cu$^{2+}$ ion as a function of temperature
for $T \leq$ 100\,K in fields $B = 0.1$~T for $B \parallel c$ and $B
\perp c$ both after cooling in zero field (ZFC). At 300~K (not shown)
$n_{\rm eff}$ is about 1.69~$\mu_{\rm B}/{\rm Cu}^{2+}$, a value
slightly smaller than the one for isolated spin-$1/2$ ions. With
decreasing temperature $n_{\rm eff}$ becomes continuously reduced down
to approximately 1.2~$\mu_{\rm B}$ around 20~K. Such a large effective
moment in an antiferromagnetically coupled system at a temperature $T
\lesssim J_3/10$ is only possible when in addition a ferromagnetic
coupling exists which is of similar size. Upon cooling to below about
18~K, $n_{\rm eff}$ for both orientations starts to increase with a
maximum slope around 15~K.  Upon further cooling, $n_{\rm eff}$ passes
over a maximum followed by a rapid decrease. This behavior is a clear
signature of a phase transition into long-range antiferromagnetic
order characterized by canted spins exhibiting a small ferromagnetic
component. This is also corroborated by the observation of a
hysteresis loop observed at 2~K and shown in
Fig.~\ref{fig:suscept}. The maximum of $n_{\rm eff}$ for the field
aligned parallel to the hexagonal planes exceeds that for fields
oriented perpendicular to the planes by nearly a factor of two,
indicating that the easy axis of barlowite is oriented along the
hexagonal plane.

Figure \ref{fig:magnetization} exhibits the magnetization measured for
$B \parallel c$ at 2\,K. The blue open squares correspond to the
virgin curve taken after cooling in zero field. This was possible
after carefully compensating for finite remanent fields present in the
superconducting magnet of the SQUID magnetometer. Also shown is the
full hysteresis loop obtained upon cycling the field in the range
$-5~{\rm T} \leq B \leq +5~{\rm T}$. The hysteresis loop closes at
$\big| B \big| \geq 4~{\rm T}$ at 2~K, cf. lower right inset to
Fig.~\ref{fig:magnetization}.  Upon field cycling (red symbols),
following the virgin run, the magnetization exhibits a large slope for
fields below 0.1~T. At this field level, $M$ reaches approximately
10\% of the expected saturation magnetization of 1~$\mu_{\rm B}$ per
Cu$^{2+}$ ion. On further increasing the field $M(B)$ increases almost
linearly and reaches a value of 0.23~$\mu_{\rm B}/{\rm Cu}^{2+}$ at
5~T, which is close to 1~$\mu_{\rm B}$/f.u. The evolution of $M$ upon
varying $B$ indicates a discontinuous magnetization process which has
to be taken into account when analyzing magnetic data taken at fields
below 4\,T. The upper inset of Fig.~\ref{fig:magnetization} shows the
low-field sector of the hysteresis loop. For $B$ = 0 barlowite has a
remanent magnetization of approximately 0.3 $\mu_{\rm B}$ per formula
unit.  According to Ref.~\onlinecite{Kahn1993} this value together
with the saturation magnetization can be used to determine the tilt
angle relative to the perfectly antiferromagnetically aligned spins
resulting in a canting angle of approximately 4.5$^\circ$. Such a
moderate spin canting out of the hexagonal plane is consistent with
the existence of a Dzyaloshinskii-Moriya (DM) interaction allowed by
symmetry in barlowite.  According to Ref.~\cite{Moriya}, the DM vector
{\bf D} lies within the mirror plane which in barlowite is
perpendicular to the kagome plane.  An estimate of the size of
$\big|{\bf D} \big|$ is given by $\big| {\bf D} \big| /J_{3} \simeq
\Delta g/g = 0.1$ which is substantial.

In summary, the structure of barlowite with perfectly aligned kagome
layers opens a new synthetic route for kagome-based structures with
the possibility of interesting phases such as ordered magnetic phases
with different ordering vectors, spin liquid, Dirac metal or even
unconventional superconductivity.  Our combined first principles
calculations with susceptibility measurements identifies the low
temperature behavior of barlowite as a canted antiferromagnet with a
canting angle of approximately 4.5$^\circ$.

{\it Acknowledgements.-} We would like to thank the Deutsche
Forschungsgemeinschaft for financial support through the collaborative
research unit TRR/SFB 49. F.S.-P. gratefully acknowledges the support
of the Alexander von Humboldt Foundation through a Humboldt Research
Fellowship. J.A.S. acknowledges support from the Independent
Research/Development program while serving at the National Science
Foundation.

\end{document}